\documentclass[aps,preprint]{revtex4}
\usepackage{amsfonts}
\usepackage{amsmath}
\usepackage{amssymb}
\usepackage{graphicx}

\input{tcilatex}

\begin{document}

\preprint{}
\title[ ]{Black hole and thin-shell wormhole solutions in
Einstein-Hoffman-Born-Infeld theory}
\author{S. Habib Mazharimousavi}
\email{habib.mazhari@emu.edu.tr}
\affiliation{Department of Physics, Eastern Mediterranean University, G. Magusa, north
Cyprus, Mersin 10, Turkey.}
\author{M. Halilsoy}
\email{mustafa.halilsoy@emu.edu.tr}
\affiliation{Department of Physics, Eastern Mediterranean University, G. Magusa, north
Cyprus, Mersin 10, Turkey.}
\author{Z. Amirabi}
\email{zahra.amirabi@emu.edu.tr}
\affiliation{Department of Physics, Eastern Mediterranean University, G. Magusa, north
Cyprus, Mersin 10, Turkey.}
\pacs{PACS number}

\begin{abstract}
We employ an old field theory model, formulated and discussed by Born,
Infeld, Hoffman and Rosen during 1930s. Our method of cutting-gluing of
spacetimes resolves the double-valuedness in the displacement vector $\vec{D}%
(\vec{E})$, pointed out by these authors. A characteristic feature of their
model is to contain a logarithmic term, and by bringing forth such a
Lagrangian anew, we aim to attract the interest of field theorists to such a
Lagrangian. We adopt the Hoffman-Born-Infeld (HBI) Lagrangian in general
relativity to construct black holes and investigate the possibility of
viable thin-shell wormholes. In particular, the stability of thin-shell
wormholes supported by normal matter in 5-dimensional
Einstein-HBI-Gauss-Bonnet gravity is highlighted.
\end{abstract}

\maketitle

\section{Introduction}

It is a well-known fact by now that non-linear electrodynamics (NED) with
various formulations has therapeutic effects on the divergent results that
arise naturally in linear Maxwell electrodynamics. The theory introduced by
Born and Infeld (BI) in 1930s \cite{1} constitutes the most prominent member
among such class of viable NED theories. Apart from the healing power of
singularities, however, drawbacks were not completely eliminated from the
theory. One such serious handicap was pointed out by Born's co-workers
shortly after the introduction of the original BI theory. This concerns the
double-valued dependence of the displacement vector $\vec{D}\left( \vec{E}%
\right) $ as a function of the electric field $\vec{E}$ \cite{2}. That is,
for the common value of $\vec{E}$ the displacement $\overrightarrow{D}$
undergoes a branching which from physical grounds was totally unacceptable.
To overcome this particular problem, Hoffman and Infeld \cite{2} and Rosen%
\cite{3}, both published successive papers on this issue. Specifically, the
model Lagrangian proposed by Hoffman and Infeld (HI) contained a logarithmic
term with remarkable consequences. It removed, for instance, the singularity
that used to arise in the Cartesian components of the $\vec{E}$. Being
unaware of this contribution by HI, and after almost 70 years, we have
rediscovered recently the ubiquitous logarithmic term of Lagrangian while in
attempt to construct a model of elementary particle in Einstein-NED theory 
\cite{4}. In our model the spacetime is divided into two regions: the inner
region consists of the Bertotti-Robinson (BR) \cite{5} spacetime while the
outer region is a Reissner-Nordstr\"{o}m (RN) type spacetime. The radius of
our particle coincides with the horizon of the RN-type black hole solution
whereas inner BR part represents a singularity-free uniform electric field
region. The two regions and the NED are glued together at the horizon on
which the appropriate boundary conditions gave not only a feasible
geometrical model of a particle but remarkably resolved also the
double-valued property of the displacement vector. In other words, with our
technique, $\vec{D}\left( \vec{E}\right) $ turns automatically into a
single-valued function.

In this paper we wish to make further use of the Hoffman-Born-Infeld (HBI)
Lagrangian in general relativity, more specifically, in constructing
4-dimensional (4D) regular black holes and thin-shell wormholes. The
wormholes in 4D requires, unfortunately, exotic matter to survive. We extend
our model also to 5D-Gauss-Bonnet (GB) theory and search for the possibility
of wormholes dominated by normal (i.e. satisfying the energy conditions)
matter rather than exotic matter. Our analysis reveal that for the negative
GB parameter $\left( \alpha <0\right) $ 5D thin-shell wormholes supported by
normal matter exists, and they are stable against linear radial
perturbations. Due to the intricate structure of the potential function,
stability analysis is carried out numerically.

Organization of the paper is as follows. In Sec. II we adopt the HBI
formalism to general relativity. Construction of regular black holes in EHBI
gravity is presented in Sec. III. Thin-shell wormholes in EHBI theory
follows in Sec. IV. In sec. V we give 5D black hole and wormholes in EHBIGB
theory with emphasis on stability analysis of wormholes given in Sec. VI. We
finalize the paper with Conclusion, which appears in Sec. VII.

\section{Review of the HBI Approach in General Relativity}

Singularity for classical charged elementary particles leads to infinite
self-electromagnetic energy. This should be removed from the Maxwell theory
of charged particles and in this regard Born and Infeld (BI) introduced a
non-linear electrodynamics such that successfully they solved the problem in
some senses \cite{1}. Briefly, we can summarize their proposal in curved
spacetime by considering a spherically symmetric pure electrically charged
particle described by the line element%
\begin{equation}
ds^{2}=-f\left( r\right) dt^{2}+\frac{1}{f\left( r\right) }%
dr^{2}+r^{2}\left( d\theta ^{2}+\sin ^{2}\theta d\varphi ^{2}\right) .
\end{equation}%
They aimed to have a non-singular electric field (we use $c=\hslash
=k_{B}=8\pi G=\frac{1}{4\pi \epsilon _{\circ }}=1$) with radial component 
\begin{eqnarray}
E_{r} &=&\frac{q}{\sqrt{q^{2}b^{2}+r^{4}}}, \\
\text{(}b &=&\text{constant, the BI parameter, and }q=\text{constant charge) 
}  \notag
\end{eqnarray}%
which means that the Maxwell $2-$form is of the form 
\begin{equation}
\mathbf{F=}E_{r}dt\wedge dr.
\end{equation}%
The corresponding action is 
\begin{equation}
S=\frac{1}{2}\int d^{4}x\sqrt{-g}\mathcal{L}\left( F,^{\star }F\right) ,
\end{equation}%
in which $F=F_{\mu \nu }F^{\mu \nu },$ $^{\star }F=F_{\mu \nu }{}^{\star
}F^{\mu \nu }$ and $^{\star }$ stands for duality (here we only consider the
static, spherically symmetric electric field such that $^{\star }F=0)$.
Easily one finds the Maxwell equation modified into 
\begin{gather}
d\left( \mathcal{L}_{F}\ {}^{\star }\mathbf{F}\right) =0 \\
\left( \mathcal{L}_{F}=\frac{\partial \mathcal{L}}{\partial F}\right)  \notag
\end{gather}%
which reveals 
\begin{equation}
d\left( \mathcal{L}_{F}\ {}E_{r}r^{2}\sin \theta d\theta \wedge d\varphi
\right) =0,
\end{equation}%
or%
\begin{equation}
\mathcal{L}_{F}\ E_{r}{}=\frac{c}{r^{2}}.
\end{equation}%
Since $F=F_{\mu \nu }F^{\mu \nu }=-2$ $E_{r}{}^{2}$ and $r^{2}=\sqrt{%
q^{2}\left( \frac{1-b^{2}E_{r}{}^{2}}{E_{r}{}^{2}}\right) }=\sqrt{%
-q^{2}\left( \frac{2+b^{2}F}{F}\right) }$it yields%
\begin{equation}
\mathcal{L}_{F}\ {}=c\sqrt{\frac{2}{2+b^{2}F}}
\end{equation}%
where $c=$constant of integration, which is identified as the charge $q$.
Solution for the Lagrangian, after adjusting the constants, takes the form of%
\begin{equation}
\mathcal{L}=\frac{4}{b^{2}}\left( 1-\sqrt{1+\frac{b^{2}F}{2}}\right) ,
\end{equation}%
i.e., the BI Lagrangian.

This example gives an idea of how simple it is to find a Lagrangian which
yields a non-singular electric field, but the question was whether this much
was enough. Hoffman and Infeld \cite{2} shortly after the BI non-linear
Lagrangian, pointed this problem out and tried to get rid of any possible
difficulties.

In \cite{2} the authors remarked that although the electric field becomes
finite at $r=0$ it yields a discontinuity, for instance in the Cartesian
component $E_{x}.$ To quote from \cite{2} "It is evident that any finite
value for $E_{r}$ at $r=0$ will lead to a discontinuity of this type".
Accordingly, their proposal alternative to the BI Lagrangian can be
summarized as follows. The simplest non-singular electric field which takes
zero value at $r=0$ can be written as%
\begin{equation}
E_{r}=\frac{qr^{2}}{\left( q^{2}b^{2}+r^{4}\right) },
\end{equation}%
so that $r^{2}$ in terms of $F$ is 
\begin{equation}
r^{2}=q\frac{1\pm \sqrt{1-4b^{2}E_{r}^{2}}}{2E_{r}}=q\frac{1\pm \sqrt{%
1+2b^{2}F}}{\sqrt{-2F}}
\end{equation}%
where $+$ and $-$ stand for $r^{4}>q^{2}b^{2}$ and $r^{4}<q^{2}b^{2},$
respectively. From (7) we find 
\begin{equation}
\mathcal{L}_{F}\ =\frac{2c}{1\pm \sqrt{1+2b^{2}F}}
\end{equation}%
where the positive branch leads to the Lagrangian%
\begin{equation}
\mathcal{L}_{+}=-\frac{2}{b^{2}}\left( k+\alpha \epsilon _{+}-\ln \epsilon
_{+}\right)
\end{equation}%
with $\alpha =1,$ $k=\ln 2-2$ and $\epsilon _{+}=1+\sqrt{1+2b^{2}F}.$ Let us
note that we wrote the Lagrangian in this form to show consistency with \cite%
{2}. Again we remind that the constant $c$ has been chosen in such a way
that $\lim_{b\rightarrow 0}\mathcal{L}_{+}=-F,$ which is the Maxwell limit.
In analogy, the negative branch gives 
\begin{equation}
\mathcal{L}_{-}=-\frac{2}{b^{2}}\left( k+\alpha \epsilon _{-}-\ln \left\vert
\epsilon _{-}\right\vert \right)
\end{equation}%
where $\epsilon _{-}=1-\sqrt{1+2b^{2}F}.$ It should be noted that, here one
does not expect the Maxwell limit as $b$ goes to zero. In fact, since $%
\mathcal{L}_{-}$ is defined for $r^{4}<q^{2}b^{2},$ automatically $b$ can
not be zero unless $r$ also goes to zero in which, the case $\mathcal{L}_{-}$
becomes meaningless.

Having $\mathcal{L}_{+}$ for $r^{4}>q^{2}b^{2}$ and $\mathcal{L}_{-}$ for $%
r^{4}<q^{2}b^{2}$ imposes $\left( \mathcal{L}_{+}=\mathcal{L}_{-}\right)
_{r^{4}=q^{2}b^{2}}$ which is satisfied, as it should. Also at $%
r^{4}=q^{2}b^{2},$ one gets $E_{r}=\frac{1}{2b}$ which is the maximum value
that $E_{r}$ may take.

Based on the criticisms made in \cite{2}, as mentioned above, we see that
this Lagrangian removes the discontinuity in, say, $E_{x}.$ So, shall we
adopt this Lagrangian for further results? The answer was given few years
later by Rosen \cite{3}, which was not affirmative. The crux of the problem
lies in the relation between $E_{r}$ and $D_{r}.$ Let us go back to the
previous case (10) once more. It is known from non-linear electrodynamics 
\cite{1,2,3} that 
\begin{equation}
D_{r}=\mathcal{L}_{F}E_{r}=\frac{q}{r^{2}}
\end{equation}%
which is singular at $r=0$. Of course, being singular for $D_{r}$ does not
matter; the problem arises once we consider $D_{r}$ as a function of $E_{r}$%
. In this way at $r=0$, $E_{r}=0$ and $D_{r}=\infty ,$ and once $r=\infty $
again $E_{r}=0,$ but $D_{r}=0$. This means that $D_{r}$ in terms of $E_{r}$
is double-valued(i.e., $D_{r}\left( E_{r}\left( r=0\right) =0\right) =\infty 
$ and $D_{r}\left( E_{r}\left( r=\infty \right) =0\right) =0$)$.$ Concerning
this objection Rosen suggested to reject this Lagrangian and instead he
recommended that the Lagrangian should be a function of the potentials. For
the detail of his work we suggest Ref. \cite{3}, but here we wish to draw
attention to a recent paper we published \cite{4} which gives a different
solution to this problem. Before we give the detail of the solution we admit
that during the time of working on \cite{4} we were not aware about this
problem, and we did not know the Hoffman-Infeld (HI) form of Lagrangian. In
certain sense, we have rediscovered a Lagrangian of 70 years old anew, from
the hard way.!

Returning to the problem, we see that in the case of the HI Lagrangian they
used two different forms for inside and outside of the typical particle in
order to keep the spacetime spherically symmetric, static Reissner-Nordstr%
\"{o}m (RN) type. This is understandable since in 1930s RN solution was one
of the best known solution whereas the Bertotti-Robinson (BR) \cite{5}
solution was yet unknown. The latter, i.e., BR, constitutes a prominent
inner substitute to (RN) as far as Einstein-Maxwell solutions are concerned,
and resolves the singularity at $r=0$, which caused HI to worry about \cite%
{2}. As we gave the detail of such a choice in Ref. \cite{4}, one can choose 
$\mathcal{L}_{+}=-\frac{2}{b^{2}}\left( k+\alpha \epsilon _{+}-\ln \epsilon
_{+}\right) $ for all regions (i.e., $r\geq \sqrt{qb}=$ the radius of our
particle, and $r\leq \sqrt{qb}).$ For outside we adopted a RN type spacetime
while for inside we had to choose a BR type spacetime. Accordingly one finds 
\begin{equation}
E_{r}=\left\{ 
\begin{tabular}{lll}
$\frac{1}{2b},$ &  & $r\leq \sqrt{qb}$ \\ 
$\frac{qr^{2}}{\left( q^{2}b^{2}+r^{4}\right) },$ &  & $r\geq \sqrt{qb}$%
\end{tabular}%
\right.
\end{equation}%
and consequently%
\begin{equation}
D_{r}=\left\{ 
\begin{tabular}{lll}
$\frac{1}{b},$ &  & $r\leq \sqrt{qb}$ \\ 
$\frac{q}{r^{2}},$ &  & $r\geq \sqrt{qb}$%
\end{tabular}%
\right.
\end{equation}%
which clearly reveals that $D_{r}$ is not a double valued function of $E_{r}$
any more$.$ We note that in matching the two spacetimes the Lanczos
energy-momentum tensor \cite{6} was employed. Let us add further that this
is not the unique choice, so that the opposite choice also is possible. That
is, a RN type spacetime for $r\leq \sqrt{qb}$ and a BR type spacetime for $%
r\geq \sqrt{qb}.$ In this latter choice the Lagrangian is $\mathcal{L}_{-}=-%
\frac{2}{b^{2}}\left( k+\alpha \epsilon _{-}-\ln \left\vert \epsilon
_{-}\right\vert \right) $ everywhere, which yields 
\begin{equation}
E_{r}=\left\{ 
\begin{tabular}{lll}
$\frac{qr^{2}}{\left( q^{2}b^{2}+r^{4}\right) },$ &  & $r\leq \sqrt{qb}$ \\ 
$\frac{1}{2b},$ &  & $r\geq \sqrt{qb}$%
\end{tabular}%
\right.
\end{equation}%
and%
\begin{equation}
D_{r}=\left\{ 
\begin{tabular}{lll}
$\frac{q}{r^{2}},$ &  & $r\leq \sqrt{qb}$ \\ 
$\frac{1}{b},$ &  & $r\geq \sqrt{qb}$%
\end{tabular}%
\right.
\end{equation}%
is again not double-valued. In Ref. \cite{4} we studied in detail the first
case alone. Obviously, the second case also can be developed into a model of
elementary particle.

\section{A different aspect of the EHBI spacetime}

Once more, we start with the EHBI Lagrangian 
\begin{equation}
\mathcal{L}=\left\{ 
\begin{tabular}{lll}
$\mathcal{L}_{-},$ &  & $r\leq \sqrt{qb}$ \\ 
$\mathcal{L}_{+},$ &  & $r\geq \sqrt{qb}$%
\end{tabular}%
\right. .
\end{equation}%
where $b$ is a free parameter such that%
\begin{equation}
\lim_{b\rightarrow 0}\mathcal{L}=\lim_{b\rightarrow 0}\mathcal{L}_{+}=-F
\end{equation}%
and 
\begin{equation}
\lim_{b\rightarrow \infty }\mathcal{L}=\lim_{b\rightarrow \infty }\mathcal{L}%
_{-}=0
\end{equation}%
which are the RN and Schwarzschild (S) limits, respectively. Our action is
chosen now as 
\begin{equation}
S=\frac{1}{2}\int d^{4}x\sqrt{-g}\left( R+\mathcal{L}\left( F\right) \right)
\end{equation}%
where $R$ is the Ricci scalar for the line element (1) and $\mathcal{L}%
\left( F\right) $ is the NED Lagrangian described hitherto. The Einstein-NED
equations are 
\begin{equation}
G_{\mu }^{\nu }=T_{\mu }^{\nu }=\frac{1}{2}\left[ \mathcal{L}\left( F\right)
\delta _{\mu }^{\nu }-4\mathcal{L}_{F}F_{\mu \lambda }F^{\nu \lambda }\right]
\end{equation}%
whereas the electromagnetic field $F_{\mu \lambda }$ satisfies (5). A
solution to the Einstein equations which gives all the correct limits is%
\begin{equation}
\begin{tabular}{l}
$f\left( r\right) =1-\frac{2m}{r}+\frac{q^{2}}{3r_{\circ }^{4}}r^{2}\ln
\left( \frac{r^{4}}{r^{4}+r_{\circ }^{4}}\right) -$ \\ 
$\frac{q^{2}\sqrt{2}}{3rr_{\circ }}\left[ \tan ^{-1}\left( \frac{\sqrt{2}r}{%
r_{\circ }}+1\right) +\tan ^{-1}\left( \frac{\sqrt{2}r}{r_{\circ }}-1\right) %
\right] -\frac{q^{2}\sqrt{2}}{6rr_{\circ }}\ln \left[ \frac{r^{2}+r_{\circ
}^{2}-\sqrt{2}rr_{\circ }}{r^{2}+r_{\circ }^{2}+\sqrt{2}rr_{\circ }}\right] +%
\frac{\sqrt{2}q^{2}\pi }{3rr_{\circ }},$%
\end{tabular}%
\text{ \ \ }
\end{equation}%
where $r_{\circ }=\sqrt{qb}$ and $m$ is the corresponding mass of S (and RN)
source. One can easily show that%
\begin{equation}
\lim_{b\rightarrow 0}f\left( r\right) =1-\frac{2m}{r}+\frac{q^{2}}{r^{2}}
\end{equation}%
and 
\begin{equation}
\lim_{b\rightarrow \infty }f\left( r\right) =1-\frac{2m}{r}.
\end{equation}%
It is interesting to observe that although the ADM mass of EHBI solution is
still $m$, the effective mass depends on charge and HBI parameter, i.e., 
\begin{equation}
m_{eff}=m-\frac{\sqrt{2}q^{2}\pi }{6r_{\circ }}.
\end{equation}%
Here one may set the effective mass to zero (note that the ADM mass of the
EHBI is not zero and survives in the metric indirectly) i.e.,%
\begin{equation}
m_{ADM}=\frac{\sqrt{2}q^{2}\pi }{6r_{\circ }}
\end{equation}%
to get a regular metric function whose Kretschmann and Ricci scalars are
finite at any point. It should be noted that this is not the case for the
regular solution mentioned in \cite{2}, i.e. in contrast to \cite{2}, our
EHBI black hole is not massless.

By employing the solution (25) now we proceed to investigate some
thermodynamical properties of the EHBI black hole. To do so we find the
horizon of the BH by equating the metric function to zero, which gives the
effective mass in terms of the horizon radius%
\begin{equation}
\begin{tabular}{l}
$m_{eff}=\frac{r_{h}}{2}\left( 1+\frac{q^{2}}{3r_{\circ }^{4}}r_{h}^{2}\ln
\left( \frac{r_{h}^{4}}{r_{h}^{4}+r_{\circ }^{4}}\right) \right. -$ \\ 
$\left. \frac{q^{2}\sqrt{2}}{3r_{h}r_{\circ }}\left[ \tan ^{-1}\left( \frac{%
\sqrt{2}r_{h}}{r_{\circ }}+1\right) +\tan ^{-1}\left( \frac{\sqrt{2}r_{h}}{%
r_{\circ }}-1\right) \right] -\frac{q^{2}\sqrt{2}}{6r_{h}r_{\circ }}\ln %
\left[ \frac{r_{h}^{2}+r_{\circ }^{2}-\sqrt{2}r_{h}r_{\circ }}{%
r_{h}^{2}+r_{\circ }^{2}+\sqrt{2}r_{h}r_{\circ }}\right] \right) ,\ \
r_{h}>r_{\circ }.$%
\end{tabular}%
\end{equation}%
Hawking temperature in terms of the event horizon radius is given
accordingly by%
\begin{equation}
T_{H}=\frac{1}{4\pi r_{h}}\left( 1-\frac{q^{2}r_{h}^{2}}{r_{\circ }^{4}}\ln
\left( 1+\frac{r_{\circ }^{4}}{r_{h}^{4}}\right) \right) .
\end{equation}%
Also the heat capacity, which is defined as%
\begin{equation}
C_{q}=T_{H}\left( \frac{\partial S\left( r\right) }{\partial T_{H}}\right)
_{q},
\end{equation}%
is given by%
\begin{equation}
C_{q}=\frac{\pi r_{h}^{2}\left( r_{h}^{4}+r_{\circ }^{4}\right) \left(
q^{2}r_{h}^{2}\ln \left( \frac{r_{h}^{4}}{r_{h}^{4}+r_{\circ }^{4}}\right)
+r_{\circ }^{4}\right) }{q^{2}r_{h}^{2}\left( r_{h}^{4}+r_{\circ
}^{4}\right) \ln \left( \frac{r_{h}^{4}}{r_{h}^{4}+r_{\circ }^{4}}\right)
-r_{\circ }^{8}+\left( 4q^{2}-r_{h}^{2}\right) r_{h}^{2}r_{\circ }^{4}}
\end{equation}%
whose zeros of the denominator indicate possible phase transitions.

\section{Thin-shell wormholes in $4D$}

Following the establishment of black hole solutions in the EHBI action (23)
with line element (1) our next venture is to investigate the possibility of
thin-shell wormholes in the same theory. Here we follow the standard method
of constructing a thin-shell wormhole \cite{7}. To do so, we take two copies
of EHBI spacetimes, and from each manifold we remove the following $4D$
submanifold%
\begin{equation}
\Omega _{1,2}\equiv \left\{ r_{1,2}\leq a\left\vert a>\sqrt{qb}\right.
\right\} 
\end{equation}%
in which $a$ is a constant and $b$ is the HBI parameter introduced before.
In addition, we restrict our free parameters to keep our metric function
non-zero and positive for $r>\sqrt{qb}$. In order to have a complete
manifold we define a manifold $\mathcal{M}=$ $\Omega _{1}\cup \Omega _{2}$
whose boundary is given by the two timelike hypersurfaces 
\begin{equation}
\partial \Omega _{1,2}\equiv \left\{ r_{1,2}=a\left\vert a>\sqrt{qb}\right.
\right\} .
\end{equation}%
After identifying the two hypersurfaces, $\partial \Omega _{1}\equiv
\partial \Omega _{2}=\partial \Omega $, the resulting manifold will be
geodesically complete \cite{7} with two asymptotically flat regions
connected by a traversable Lorantzian wormhole. The throat of the wormhole
is at $\partial \Omega $ and the induced metric on $\mathcal{M}$ with
coordinate $\left\{ X^{i}\right\} $ and induced metric $h_{ij},$ takes the
form \ 
\begin{equation}
ds_{ind}^{2}=-d\tau ^{2}+a\left( \tau \right) ^{2}\left( d\theta ^{2}+\sin
^{2}\theta d\phi ^{2}\right) 
\end{equation}%
where $\tau $ represents the proper time on the hypersurface $\partial
\Omega .$ Lanczos equations \cite{7} read%
\begin{equation}
S_{j}^{i}=-\frac{1}{8\pi }\left( \left[ K_{j}^{i}\right] -\delta _{j}^{i}%
\left[ K\right] \right) ,
\end{equation}%
where the extrinsic curvature $K_{ij}$ (with trace $K$) is defined by 
\begin{equation}
K_{ij}=-n_{k}\left( \frac{\partial ^{2}X^{k}}{\partial \xi ^{i}\partial \xi
^{j}}+\Gamma _{mn}^{k}\frac{\partial X^{m}}{\partial \xi ^{i}}\frac{\partial
X^{n}}{\partial \xi ^{j}}\right) ,
\end{equation}%
in which $n_{k}$ is normal to $\mathcal{M}$, so that $%
h_{ij}=g_{ij}-n_{i}n_{j}$ and $\xi ^{i}=\left( \tau ,\theta ,\phi \right) .$
Upon substitution into (37) we obtain the surface stress-energy tensor in
the form 
\begin{equation}
S_{j}^{i}=diag\left( -\sigma ,p_{\theta },p_{\phi }\right) .
\end{equation}%
Here $\sigma ,$ and $p_{\theta \text{ }}=p_{\phi }$ are the surface-energy
density and the surface pressures, respectively. A detailed study shows \cite%
{8} that 
\begin{equation}
\sigma =-\frac{1}{2\pi a}\sqrt{f\left( a\right) +\dot{a}^{2}}
\end{equation}%
and 
\begin{equation}
p_{\theta \text{ }}=p_{\phi }=-\frac{1}{2}\sigma +\frac{1}{8\pi }\frac{2%
\ddot{a}+f^{\prime }\left( a\right) }{\sqrt{f\left( a\right) +\dot{a}^{2}}}.
\end{equation}%
Also the conservation equation gives%
\begin{equation}
\frac{d}{d\tau }\left( \sigma a^{2}\right) +p\frac{d}{d\tau }\left(
a^{2}\right) =0
\end{equation}%
or%
\begin{equation}
\dot{\sigma}+2\frac{\dot{a}}{a}\left( p+\sigma \right) =0.
\end{equation}%
For the static structure, one gets 
\begin{equation}
\sigma _{0}=-\frac{1}{2\pi a_{0}}\sqrt{f\left( a_{0}\right) },\text{ \ \ }%
p_{0}=\frac{\sqrt{f\left( a_{0}\right) }}{4\pi a_{0}}\left( 1+\frac{a}{2}%
\frac{f^{\prime }\left( a_{0}\right) }{f\left( a_{0}\right) }\right) .
\end{equation}%
The total amount of exotic matter for constructing such a thin-shell
wormhole is given by%
\begin{equation}
\Omega =\tint \left( \rho +p\right) \sqrt{-g}d^{3}x.
\end{equation}%
Here $\rho =\delta \left( r-a\right) \sigma \left( a\right) $ where $\delta
\left( .\right) $ is the Dirac delta function, radial pressure $p$ is
negligible because of the thin shell, and therefore 
\begin{equation}
\Omega =4\pi a^{2}\sigma \left( a\right) =-2a\sqrt{f\left( a\right) }.
\end{equation}%
The EHBI metric function now takes the form%
\begin{equation}
\begin{tabular}{l}
$f\left( r\right) =1-\frac{2m}{r}+\frac{q^{2}}{3r_{\circ }^{4}}r^{2}\ln
\left( \frac{r^{4}}{r^{4}+r_{\circ }^{4}}\right) -$ \\ 
$\frac{q^{2}\sqrt{2}}{3rr_{\circ }}\left[ \tan ^{-1}\left( \frac{\sqrt{2}r}{%
r_{\circ }}+1\right) +\tan ^{-1}\left( \frac{\sqrt{2}r}{r_{\circ }}-1\right) %
\right] -\frac{q^{2}\sqrt{2}}{6rr_{\circ }}\ln \left[ \frac{r^{2}+r_{\circ
}^{2}-\sqrt{2}rr_{\circ }}{r^{2}+r_{\circ }^{2}+\sqrt{2}rr_{\circ }}\right] +%
\frac{\sqrt{2}q^{2}\pi }{3rr_{\circ }},\ \ r>r_{\circ },$%
\end{tabular}%
\end{equation}%
in which $r_{\circ }=\sqrt{qb}.$\ The equation of motion for the thin-shell
wormhole can be extracted from (40) 
\begin{equation}
\dot{a}^{2}+V\left( a\right) =0,
\end{equation}%
in which the thin shell's potential is given by%
\begin{equation}
V\left( a\right) =f\left( a\right) -\left( 2\pi a\sigma \left( a\right)
\right) ^{2}.
\end{equation}%
In order to investigate the radial perturbation around an equilibrium radius
($a_{0}$) we assume a linear relation between the pressure and density%
\begin{equation}
p=p_{0}+\beta ^{2}\left( \sigma -\sigma _{0}\right) ,
\end{equation}%
in which $p_{0}$, $\sigma _{0}$ and $\beta $ are constants. Upon expansion
around $a_{0}$ (which requires $V\left( a_{0}\right) =$ $V^{\prime }\left(
a_{0}\right) =0$) up to the second order yields%
\begin{equation}
V\left( a\right) \cong \frac{1}{2}V^{\prime \prime }\left( a_{0}\right)
\left( a-a_{0}\right) ^{2}.
\end{equation}%
By considering (43) and using $\sigma ^{\prime }=\frac{\dot{\sigma}}{\dot{a}}
$ one gets%
\begin{equation}
V^{\prime \prime }\left( a_{0}\right) =f_{0}^{\prime \prime }-\frac{%
f_{0}^{\prime 2}}{2f_{0}}-\frac{1+2\beta ^{2}}{a_{0}^{2}}\left(
2f_{0}-a_{0}f_{0}^{\prime }\right) 
\end{equation}%
in which  $f_{0}=f\left( a_{0}\right) .$ The stability conditions $V^{\prime
\prime }\left( a_{0}\right) \geq 0$ leads to%
\begin{equation}
\text{for }2f_{0}\gtrless a_{0}f_{0}^{\prime }\text{, \ \ \ \ \ \ \ \ \ \ }%
1+2\beta ^{2}\lessgtr \frac{a_{0}^{2}}{2f_{0}}\left( \frac{2f_{0}^{\prime
\prime }f_{0}-f_{0}^{\prime 2}}{2f_{0}-a_{0}f_{0}^{\prime }}\right) .
\end{equation}%
Since its source is already exotic matter we shall not be interested in this
particular wormhole any further in the present paper. Instead, we shall go
to 5D, in which the black hole and wormhole constructions render it possible
to make normal matter, stable wormholes. This is the main strategy in the
following chapters.

\section{$5-$dimensional EHBI black hole}

In order to extend the $4D$ EHBI black hole solution to $5D$ with a
cosmological constant $\Lambda $ we choose our action as%
\begin{equation}
S=\frac{1}{2}\int dx^{5}\sqrt{-g}\left\{ -4\Lambda +R+\mathcal{L}\left( 
\mathcal{F}\right) \right\} ,
\end{equation}%
where%
\begin{equation}
\mathcal{L}=\left\{ 
\begin{tabular}{lll}
$\mathcal{L}_{-},$ &  & $r\leq \sqrt{qb}$ \\ 
$\mathcal{L}_{+},$ &  & $r\geq \sqrt{qb}$%
\end{tabular}%
\right. 
\end{equation}%
and the nonlinear Maxwell equation (5) in 5D leads to the radial electric
field%
\begin{equation}
E_{r}=\frac{qr^{3}}{\left( q^{2}b^{2}+r^{6}\right) }.
\end{equation}%
Variation of the action (54) yields the field equations as 
\begin{gather}
G_{\mu }^{\ \nu }+2\Lambda \delta _{\mu }^{\ \nu }=T_{\mu }^{\ \nu }, \\
T_{\mu }^{\ \nu }=\frac{1}{2}\left( \mathcal{L}\delta _{\mu }^{\ \nu }-4%
\mathcal{L}_{\mathcal{F}}F_{\mu \lambda }F^{\nu \lambda }\right) ,  \notag
\end{gather}%
which clearly gives $T_{t}^{\ t}=T_{r}^{\ r}=\left( \frac{1}{2}\mathcal{L}-%
\mathcal{L}_{\mathcal{F}}\mathcal{F}\right) ,$ stating also that $G_{t}^{\
t}=G_{r}^{\ r}$ and $T_{\theta _{i}}^{\ \theta _{i}}=\frac{1}{2}\mathcal{L}$%
. Now, we introduce our ansatz line element ($\chi =\pm 1,0$)%
\begin{equation}
ds^{2}=-(\chi -r^{2}H\left( r\right) )dt^{2}+\frac{1}{(\chi -r^{2}H\left(
r\right) )}dr^{2}+r^{2}d\Omega _{3}^{2}
\end{equation}%
in which $H\left( r\right) $ is a function to be determined, to cover both
the topological and non-topological black hole solutions \cite{9}. Our
choice of $g_{tt}=-\left( g_{rr}\right) ^{-1}$ is a direct result of $%
G_{t}^{\ t}=G_{r}^{\ r}$ up to a constant coefficient, which is chosen to be
one. The Einstein tensor components are given by%
\begin{eqnarray}
G_{t}^{\ t} &=&G_{r}^{\ r}=-\frac{3}{2r^{3}}\left( r^{4}H\left( r\right)
\right) ^{\prime }  \notag \\
G_{\theta _{i}}^{\ \theta _{i}} &=&-\frac{1}{2r^{2}}\left( r^{4}H\left(
r\right) \right) ^{\prime \prime }
\end{eqnarray}%
from which, one obtains a general class of $H\left( r\right) $ functions
depending on the choice of $T_{t}^{t}$, 
\begin{equation}
H\left( r\right) =\frac{\Lambda }{3}+\frac{4m}{\left( d-2\right) r^{d-1}}-%
\frac{2}{\left( d-2\right) r^{d-1}}\tint r^{d-2}T_{t}^{t}dr.
\end{equation}%
Now, with the particular choice of the energy-momentum tensor component as 
\begin{equation}
T_{t}^{t}=-\frac{1}{b^{2}}\ln \left( 1+\frac{b^{2}q^{2}}{r^{6}}\right) 
\end{equation}%
the metric function is found to be 
\begin{equation}
\begin{tabular}{l}
$f\left( r\right) =\chi -\frac{\Lambda }{3}r^{2}-\frac{4m}{3r^{2}}-\frac{%
q^{2}\sqrt{3}}{6r^{2}r_{\circ }^{2}}\tan ^{-1}\left( \frac{1}{\sqrt{3}}\left[
\frac{2r^{2}}{r_{\circ }^{2}}-1\right] \right) -$ \\ 
$\frac{q^{2}}{12r^{2}r_{\circ }^{2}}\ln \left\vert \frac{r^{4}+r_{\circ
}^{4}-r^{2}r_{\circ }^{2}}{r^{4}+r_{\circ }^{4}+2r^{2}r_{\circ }^{2}}%
\right\vert +\frac{1}{6}\frac{r^{2}q^{2}}{r_{\circ }^{6}}\ln \left( \frac{%
r^{6}}{r^{6}+r_{\circ }^{6}}\right) +\frac{\sqrt{3}q^{2}\pi }{%
12r^{2}r_{\circ }^{2}},$%
\end{tabular}%
,
\end{equation}%
where $r_{\circ }^{6}=b^{2}q^{2}$ and $m$ is the ADM mass of the black hole.
One observes that this solution in two extremal limits for $b$ yields%
\begin{eqnarray}
\lim_{b\rightarrow 0}f\left( r\right)  &=&\chi -\frac{\Lambda }{3}r^{2}-%
\frac{4m}{3r^{2}}+\frac{q^{2}}{3r^{4}},  \notag \\
\lim_{b\rightarrow \infty }f\left( r\right)  &=&\chi -\frac{\Lambda }{3}%
r^{2}-\frac{4m}{3r^{2}}.
\end{eqnarray}%
Further, in the sense of usual ADM mass, even if one adjusts 
\begin{equation}
m_{ADM}=\frac{\sqrt{3}q^{2}\pi }{16r_{\circ }^{2}}
\end{equation}%
unlike the case of 4D, the metric remains singular at origin. 

\section{$5D$ stable, normal matter thin-shell wormhole in EHBIGB theory}

Our action and metric ansatz in 5D EHBIGB theory of gravity are given
respectively by%
\begin{equation}
S=\frac{1}{2}\int dx^{5}\sqrt{-g}\left\{ -4\Lambda +R+\alpha \tciLaplace
_{GB}+\mathcal{L}\left( \mathcal{F}\right) \right\} 
\end{equation}%
and 
\begin{equation}
ds^{2}=-f\left( r\right) dt^{2}+\frac{1}{f\left( r\right) }%
dr^{2}+r^{2}\left( d\theta ^{2}+\sin ^{2}\theta \left( d\phi ^{2}+\sin
^{2}\phi d\psi ^{2}\right) \right) 
\end{equation}%
where $\tciLaplace _{GB}=R_{\mu \nu \gamma \delta }R^{\mu \nu \gamma \delta
}-4R_{\mu \nu }R^{\mu \nu }+R^{2}$ and $\alpha $ is the GB parameter. The
inclusion of the GB term modifies (57) and (58), which can be expressed as
an algebraic equation for $H\left( r\right) $ given by%
\begin{equation}
H\left( r\right) +4\alpha H\left( r\right) ^{2}=\frac{\Lambda }{3}+\frac{4m}{%
3r^{4}}-\frac{2}{3r^{4}}\tint r^{3}T_{t}^{t}dr.
\end{equation}%
Upon insertion of (61) for $T_{t}^{t}$ one obtains $H\left( r\right) ,$ and
as a result%
\begin{equation}
\begin{tabular}{l}
$f_{\pm }\left( r\right) =\chi +\frac{r^{2}}{8\alpha }\times \left\{ 1\pm 
\sqrt{1+16\alpha H}\right\} ,$ \\ 
$H=\frac{\Lambda }{3}+\frac{4m_{eff}}{3r^{4}}+\frac{q^{2}\sqrt{3}}{%
6r^{4}r_{\circ }^{2}}\tan ^{-1}\left( \frac{1}{\sqrt{3}}\left[ \frac{2r^{2}}{%
r_{\circ }^{2}}-1\right] \right) +\frac{q^{2}}{12r^{4}r_{\circ }^{2}}\ln
\left\vert \frac{r^{4}+r_{\circ }^{4}-r^{2}r_{\circ }^{2}}{r^{4}+r_{\circ
}^{4}+2r^{2}r_{\circ }^{2}}\right\vert -\frac{1}{6}\frac{q^{2}}{r_{\circ
}^{6}}\ln \left( \frac{r^{6}}{r^{6}+r_{\circ }^{6}}\right) $%
\end{tabular}%
\end{equation}%
in which $r_{\circ }^{6}=b^{2}q^{2}$ and $m_{eff}=m-\frac{\sqrt{3}q^{2}\pi }{%
16r_{\circ }^{2}}.$ One can easily check the following limits%
\begin{eqnarray}
&&%
\begin{tabular}{l}
$\lim_{b\rightarrow 0}f_{\pm }\left( r\right) =\chi +\frac{r^{2}}{8\alpha }%
\left\{ 1\pm \sqrt{1+16\alpha \left( \frac{\Lambda }{3}+\frac{4m}{3r^{4}}-%
\frac{q^{2}}{3r^{6}}\right) }\right\} ,$%
\end{tabular}
\\
&&%
\begin{tabular}{l}
$\lim_{b\rightarrow \infty }f_{\pm }\left( r\right) =\chi +\frac{r^{2}}{%
8\alpha }\left\{ 1\pm \sqrt{1+16\alpha \left( \frac{\Lambda }{3}+\frac{4m}{%
3r^{4}}\right) }\right\} ,$%
\end{tabular}
\notag \\
&&%
\begin{tabular}{l}
$\lim_{\alpha \rightarrow 0}f_{-}\left( r\right) =\chi -\frac{\Lambda }{3}%
r^{2}-\frac{4m_{eff}}{3r^{2}}-\frac{q^{2}\sqrt{3}}{6r^{2}r_{\circ }^{2}}\tan
^{-1}\left( \frac{1}{\sqrt{3}}\left[ \frac{2r^{2}}{r_{\circ }^{2}}-1\right]
\right) -$ \\ 
$\frac{q^{2}}{12r^{2}r_{\circ }^{2}}\ln \left\vert \frac{r^{4}+r_{\circ
}^{4}-r^{2}r_{\circ }^{2}}{r^{4}+r_{\circ }^{4}+2r^{2}r_{\circ }^{2}}%
\right\vert +\frac{1}{6}\frac{r^{2}q^{2}}{r_{\circ }^{6}}\ln \left( \frac{%
r^{6}}{r^{6}+r_{\circ }^{6}}\right) ,$%
\end{tabular}
\notag
\end{eqnarray}%
as expected. With the solution (68), (66) represents a 5D black hole in
EHBIGB gravity and now we shall proceed to construct a thin-shell wormhole
solution in the same spacetime. For this process it is necessary to remove
the regions 
\begin{equation}
M_{1,2}=\left\{ r_{1,2}\leq a,\text{ \ \ }a>r_{h}\right\} 
\end{equation}%
from the underlying spacetime. Here $r_{h}$ is the event horizon and
subsequently we paste the remaining regions of spacetime to provide geodesic
completeness. The time-like boundary surface $\Sigma _{1,2}$ on $M_{1,2}$
are glued such that%
\begin{equation}
\Sigma _{1,2}=\left\{ r_{1,2}=a,\text{ \ \ }a>r_{h}\right\} .
\end{equation}%
From the Darmois-Israel formalism \cite{10}, in terms of the original
coordinates $x^{\gamma }=\left( t,r,\theta ,\phi ,\psi \right) ,$ we define
the new set of coordinates $\xi ^{i}=\left( \tau ,\theta ,\phi ,\psi \right)
,$ with $\tau $ the proper time. Following the generalized Darmois-Israel
junction conditions apt for the GB gravity \cite{11} a surface
energy-momentum tensor is defined by $S_{j}^{i}=diag\left( \sigma ,p_{\theta
},p_{\phi },p_{\psi }\right) $, which has already been defined in terms of
the extrinsic curvature of induced metric in 4D in Sec. IV. By employing
this formalism, Richarte and Simeone \cite{12} established a thin shell
wormhole in EMGB gravity supported by normal matter. The thin-shell geometry
whose radius is assumed a function of proper time is given by%
\begin{equation}
\Sigma :f\left( r,\tau \right) =r-a\left( \tau \right) =0.
\end{equation}%
The generalized Darmois-Israel conditions on $\Sigma $ determines the
surface energy-momentum tensor. $S_{ab}$ which is expressed by \cite{11} 
\begin{equation}
-\frac{1}{8}S_{ab}=2\left\langle K_{ab}-Kh_{ab}\right\rangle +4\alpha
\left\langle 3J_{ab}-Jh_{ab}+2P_{acdb}K^{cd}\right\rangle .
\end{equation}%
Here a bracket implies a jump across $\Sigma $, and $h_{ab}=g_{ab}-n_{a}n_{b}
$ is the induced metric with coordinate set $\left\{ X^{a}\right\} $ which
helps to define the extrinsic curvature introduced in Sec. IV. The
divergence-free part of the Riemann tensor $P_{abcd}$ and the tensor $J_{ab}$
(with trace $J=J_{a}^{a}$) are given by \cite{11}%
\begin{align}
P_{abcd}& =R_{abcd}+\left( R_{bc}h_{da}-R_{bd}h_{ca}\right) -\left(
R_{ac}h_{db}-R_{ad}h_{cb}\right) +\frac{1}{2}R\left(
h_{ac}h_{db}-h_{ad}h_{cb}\right) , \\
J_{ab}& =\frac{1}{3}\left[
2KK_{ac}K_{b}^{c}+K_{cd}K^{cd}K_{ab}-2K_{ac}K^{cd}K_{ab}-K^{2}K_{ab}\right] .
\end{align}%
By employing these expressions through (73) we find the energy density and
surface pressures for a generic metric function $f\left( r\right) ,$ with $%
r=a\left( \tau \right) .$ The results are given by \cite{13} 
\begin{eqnarray}
\sigma  &=&-S_{\tau }^{\tau }=-\frac{\Delta }{4\pi }\left[ \frac{3}{a}-\frac{%
4\alpha }{a^{3}}\left( \Delta ^{2}-3\left( 1+\dot{a}^{2}\right) \right) %
\right] , \\
S_{\theta }^{\theta } &=&S_{\phi }^{\phi }=S_{\psi }^{\psi }=p=\frac{1}{4\pi 
}\left[ \frac{2\Delta }{a}+\frac{\ell }{\Delta }-\frac{4\alpha }{a^{2}}%
\left( \ell \Delta -\frac{\ell }{\Delta }\left( 1+\dot{a}^{2}\right) -2\ddot{%
a}\Delta \right) \right] ,
\end{eqnarray}%
where $\ell =\ddot{a}+f^{\prime }\left( a\right) /2$ and $\Delta =\sqrt{%
f\left( a\right) +\dot{a}^{2}}$ in which 
\begin{equation}
f\left( a\right) =\left. f_{-}\left( r\right) \right\vert _{r=a}.
\end{equation}%
We note that in our notation a 'dot' denotes derivative with respect to the
proper time $\tau $ and a 'prime' with respect to the argument of the
function. For simplicity, we set the cosmological constant to zero. It can
be checked that by substitution from (76) and (77) the conservation equation 
\begin{equation}
\frac{d}{d\tau }\left( \sigma a^{3}\right) +p\frac{d}{d\tau }\left(
a^{3}\right) =0.
\end{equation}%
holds true. For the static configuration of radius $a_{0}$ we have the
constant values 
\begin{eqnarray}
\sigma _{0} &=&-\frac{\sqrt{f\left( a_{0}\right) }}{4\pi }\left[ \frac{3}{%
a_{0}}-\frac{4\alpha }{a_{0}^{3}}\left( f\left( a_{0}\right) -3\right) %
\right] , \\
p_{0} &=&\frac{\sqrt{f\left( a_{0}\right) }}{4\pi }\left[ \frac{2}{a_{0}}+%
\frac{f^{\prime }\left( a_{0}\right) }{2f\left( a_{0}\right) }-\frac{2\alpha 
}{a_{0}^{2}}\frac{f^{\prime }\left( a_{0}\right) }{f\left( a_{0}\right) }%
\left( f\left( a_{0}\right) -1\right) \right] .
\end{eqnarray}

In order to investigate the radial perturbation around an equilibrium radius
($a_{0}$) we assume a linear relation between the pressure and density \cite%
{13}, as in the 4D case 
\begin{equation}
p=p_{0}+\beta ^{2}\left( \sigma -\sigma _{0}\right) .
\end{equation}%
Here the constant $\sigma _{0}$ and $p_{0}$ are given by (80) and (81)
whereas $\beta ^{2}$ is a constant parameter which can be identified with
the speed of sound. By virtue of the latter equation we express the energy
density in the form 
\begin{equation}
\sigma \left( a\right) =\left( \frac{\sigma _{0+}p_{0}}{\beta ^{2}+1}\right)
\left( \frac{a_{0}}{a}\right) ^{3\left( \beta ^{2}+1\right) }+\frac{\beta
^{2}\sigma _{0-}p_{0}}{\beta ^{2}+1}.
\end{equation}%
This, together with (79) lead us to the equation of motion for the radius of
throat, which reads%
\begin{equation}
-\frac{\sqrt{f\left( a\right) +\dot{a}^{2}}}{4\pi }\left[ \frac{3}{a}-\frac{%
4\alpha }{a^{3}}\left( f\left( a\right) -3-2\dot{a}^{2}\right) \right]
=\left( \frac{\sigma _{0+}p_{0}}{\beta ^{2}+1}\right) \left( \frac{a_{0}}{a}%
\right) ^{3\left( \beta ^{2}+1\right) }+\frac{\beta ^{2}\sigma _{0-}p_{0}}{%
\beta ^{2}+1}.
\end{equation}%
After some manipulation this can be cast into 
\begin{equation}
\dot{a}^{2}+V\left( a\right) =0,
\end{equation}%
where 
\begin{equation}
V\left( a\right) =f\left( a\right) -\left( \left[ \sqrt{A^{2}+B^{3}}-A\right]
^{1/3}-\frac{B}{\left[ \sqrt{A^{2}+B^{3}}-A\right] ^{1/3}}\right) ^{2}
\end{equation}%
involves the root of a third order algebraic equation with 
\begin{eqnarray}
A &=&\frac{\pi a^{3}}{4\alpha }\left[ \left( \frac{\sigma _{0+}p_{0}}{\beta
^{2}+1}\right) \left( \frac{a_{0}}{a}\right) ^{3\left( \beta ^{2}+1\right) }+%
\frac{\beta ^{2}\sigma _{0-}p_{0}}{\beta ^{2}+1}\right] , \\
B &=&\frac{a^{2}}{8\alpha }+\frac{1-f\left( a\right) }{2}.
\end{eqnarray}%
It is a simple exercise to show that $V\left( a\right) ,$ and $V^{\prime
}\left( a\right) ,$ both vanish at $a=a_{0}.$ The stability requirement for
the wormhole reduces to the determination of $\ V^{\prime \prime }(a_{0})>0.$
Due to the complicated structure of the potential we shall proceed through
numerical analysis in order to explore the stability regions, if there is
any at all. In doing this, we shall concentrate mainly on the case $\alpha
<0 $, since after all, the case $\alpha >0$ does not have a good record in
the context of normal matter thin-shell wormholes.

Fig (1) displays the $V^{\prime \prime }\left( a\right) >0$ as stability
region upon projection in the two-dimensional variables $\beta $ and $a_{0}$
with $\alpha <0.$ The metric function $f\left( r\right) $ and the reality of 
$\sigma >0$ are also visible in the inscribed plots. Beyond these regions
and for the associated variables, in particular for the choice $\alpha >0$,
stable wormhole construction in EHBIGB theory doesn't seem possible.

\section{Conclusion}

The original non-linear BI electrodynamics aimed at removing point-like
singularities and resulting divergences. This, however, didn't resolve the
dauble-valuedness in the displacement vector $\vec{D}(\vec{E})$ as a
function of the electric field. This was the main motivation for emergence
of Hoffman's version of the BI type Lagrangian, which contained an
ubiquitous logarithmic term \cite{14}. We have shown that such a
supplementary term in the Lagrangian has benefits also when employed in
general relativity. Firstly, it removes the double valuedness in $\vec{D}(%
\vec{E}),$ as observed / proposed seven decades before. Secondly, by cutting
and gluing (pasting) method we construct black hole spacetime. This may be
developed into a finite, geometrical model of elementary particle as
addressed in \cite{4}. Lastly, as we have emphasized in the present paper,
the HBI type Lagrangian can be used in wormhole construction. These
wormholes have the attractive features of being supported only by normal
matter. By exploiting the boundary conditions while gluing the inner and
outer parts we remove the divergent part arising from the solution in 4D
which, however, doesn't seem possible in the Gauss-Bonnet augmented
Lagrangian in higher dimensions. Further, the thin-shell wormhole obtained
in the EHBIGB gravity the wormhole can be made stable \cite{15}. This is
upon finely-tuned parameters and an intricate potential function which is
required to have positive second derivative. From these feats it is hoped
that the logarithmic Lagrangian will draw attention from various circles of
field theorists for further applications.

\textbf{Figure caption:}

Fig. 1: The stability region (i.e. $V^{\prime \prime }(a)>0$) for the chosen
parameters, $r_{0}=1.00,$ $q=0.75$ and $m_{eff}=0$ (Eq. (28))$.$ This is
given as a projection into the plane with axes $\beta $ and $\frac{a_{0}}{%
\left\vert \alpha \right\vert }.$ The plot of the metric function $f\left(
r\right) $ and energy density $\sigma $ are also inscribed in the figure.

\bigskip

\bigskip

\end{document}